\documentclass[12pt]{myiopart}
\usepackage{amssymb,amsmath}
\usepackage{color} 
\usepackage{accents} 
\usepackage{texdraw}
\usepackage{eucal}
\usepackage{epic,epsfig}

\def\m@th{\mathsurround=0pt}
\mathchardef\bracell="0365 
\def\upbrall{$\m@th\bracell$}
\def\undertilde#1{\mathop{\vtop{\ialign{##\crcr
    $\hfil\displaystyle{#1}\hfil$\crcr
     \noalign
     {\kern1.5pt\nointerlineskip}
     \upbrall\crcr\noalign{\kern1pt
   }}}}\limits}
\def\theequation{\arabic{section}.\arabic{equation}}

\newcommand{\ssp}{\mathfrak{p}}

\newcommand{\ssq}{\mathfrak{q}}

\newcommand{\ecl}{\end{color}}

\newcommand{\dd}{\delta}
\newcommand{\sg}{\sigma}
\newcommand{\dsg}{\dot{\sigma}}
\newcommand{\dtau}{\dot{\tau}}

\newcommand{\be}{\begin{equation}}
\newcommand{\ee}{\end{equation}}
\newcommand{\bea}{\begin{eqnarray}}
\newcommand{\eea}{\end{eqnarray}}
\newcommand{\bse}{\begin{subequations}}
\newcommand{\ese}{\end{subequations}}
\newcommand{\nn}{\nonumber}

\newcommand{\wh}{\widehat}
\newcommand{\ol}{\overline}
\newcommand{\wt}{\widetilde}

\newcommand{\po}{\accentset{o}{p}}
\newcommand{\qo}{\accentset{o}{q}}

\newcommand{\bphi}{{\boldsymbol \phi}}

\date{today}

\begin{document}
\title{Soliton solutions for Q3}

\author{James Atkinson$^1$, Jarmo Hietarinta$^2$ and Frank
  Nijhoff$^1$} 

\address{ $^1$ Department of Applied Mathematics, University of Leeds,
  Leeds LS2 9JT, UK \\ 
  $^2$ Department of Physics, University of Turku, FIN-20014 Turku,
  FINLAND
}
\begin{abstract}
  We construct N-soliton solutions to the equation called Q3 in the
  recent Adler-Bobenko-Suris classification.  An essential ingredient
  in the construction is the relationship of ${\rm (Q3)}_{\dd=0}$ to the
  equation proposed by Nijhoff, Quispel and Capel in 1983 (the NQC
  equation).  This latter equation has two extra parameters, and
  depending on their sign choices we get a 4-to-1 relationship from
  NQC to ${\rm (Q3)}_{\dd=0}$. This leads to a four-term background
  solution, and then to a 1-soliton solution using a B\"acklund
  transformation.  Using the 1SS as a guide allows us to get the
  N-soliton solution in terms of the tau-function of the Hirota-Miwa
  equation.
\end{abstract}

\section{Introduction}
\setcounter{equation}{0} Integrable lattice equations have a long
history, going back to pioneering work in the 1970's, \cite{AL,Hir},
with subsequent development of systematic approaches in the early
1980's, \cite{NQC,QNCL,DJM}, cf.~also the review \cite{NC}. It has
long been known, in fact it is implicit in these constructions, that
the emerging examples possess the property of ``multidimensional
consistency''. By this we mean the property that the partial
difference equation describing the lattice systems can be consistently
embedded in a multidimensional space, with in principle an infinity of
lattice variables. Most succinctly this property was described in
recent years in \cite{NW}, in a context where it was explicitly used
to achieve multi-dimensional reductions, and it was subsequently
re-appraised in \cite{BS}. This multidimensional consistency is a very
natural property, it is the precise analogue of the well known
existence of compatible higher time-flows in hierarchies of soliton
systems, and as such it was quite well understood in the earlier
publications mentioned.

In \cite{ABS} the property of ``consistency around a cube'' (CAC) was
used to classify partial difference equations defined on an elementary
square of a two-dimensional lattice.  Remarkably, within certain
restrictive conditions (symmetry and the ``tetrahedron condition''), a
full list of scalar quadrilateral lattice equations could be
established, and this list is surprisingly short. (In a more recent paper
\cite{ABS2} the classification statement was proven under slightly
less restrictive conditions.)

The list of CAC-integrable lattice equations in \cite{ABS} contains
some interesting examples, but the equation at the top of the list
(denoted as Q4 in \cite{ABS}) was actually found earlier by V. Adler,
cf.~\cite{Adler}. This equation, which we refer to as \textit{Adler's
  equation}, is in fact an integrable discretization of the famous
Krichever-Novikov (KN) equation, \cite{KN1,KN2}.  For Adler's equation
a Lax pair was established in \cite{Nij} on the basis of its
multidimensional consistency, and in \cite{Q4} this equation was shown
to be a master equation among various well-known integrable systems
associated with an elliptic curve. The first solutions for Adler's
equation were established in a recent paper \cite{AHN}, cf.  also
\cite{AN} for a slight generalisation of those results.  However, so
far little is known about the algebraic structure behind Adler's
equation, and since this equation has lattice parameters which lie on
an elliptic curve, the underlying structure is expected to be
interesting but rather complicated. Thus, before tackling the Adler's
equation, it is of interest to study some of the other examples
emerging from the list of \cite{ABS}, in order to see how the
underlying structure of such equations can be revealed.

Here we focus on the equation denoted as Q3 in \cite{ABS} and which
is just below Q4 in the hierarchy. It is written as: 
\begin{align}\label{eq:Q3}
   \po(1-\qo^2)(u\wh{u}+\wt{u}\wh{\wt{u}})
-\qo(1-\po^2)(u\wt{u}+&\wh{u}\wh{\wt{u}})
-(\po^2-\qo^2)(\wh{u}\wt{u}+u\wh{\wt{u}}) \nn \\
   =&
  \dd^2(\po^2-\qo^2)(1-\po^2)(1-\qo^2)/(4\po\qo).
\end{align} 
In (\ref{eq:Q3}) we have adopted the convention of representing shifts
in the rectangular lattice with tildes and hats: The corners of an
elementary plaquette on a rectangular lattice are thus
\[ 
u\equiv u_{n,m},\quad \wt{u}\equiv u_{n+1,m},\quad \wh{u}\equiv u_{n,m+1},\quad
\wh{\wt{u}}\equiv u_{n+1,m+1}.
\]
The \textit{lattice parameters} $\po,\,\qo$ in \eqref{eq:Q3} are
associated with the $n,\,m$ directions in the lattice, respectively,
(they can be thought of as measuring the grid size in these
directions) while $\dd$ is a global parameter.

In this paper we construct a general family of $N$-soliton solutions
of Q3. The construction of these solutions is based on the
relationship of \eqref{eq:Q3} to a lattice equation that first
appeared in \cite{NQC} (cf.~also \cite{QNCL}) about 25 years ago, and
which is equivalent to the $\delta=0$ case of Q3.  The explicit 4-to-1
relationship between these two equations is explained in Section
\ref{S:eqs}, and it requires the introduction of another
parametrisation which is more suitable for the solution.
The N-soliton solution of Q3 with parameter $\delta$ (denoted by
$({\rm Q3})_\delta$) then appears as a {\em linear} combination of
four independent solutions of the $\delta=0$ case, $({\rm Q3})_0$,
with four arbitrary coefficients subject to one single relation.  We
have obtained this result in two different ways, one of which employs
a novel Miura transformation (explained in the Appendix), which
allows us to derive soliton solutions for $({\rm Q3})_\delta$ from the
known solutions of $({\rm Q3})_0$ of \cite{NQC,QNCL}. We will not
present this approach here, because it requires quite a bit of a
notational machinery for which we lack space in this note, and this
approach will be published elsewhere, \cite{AN2}.  

Alternatively, we can use the 4-to-1 correspondence which appears
through the introduction of some additional lattice shifts associated
with the parametrisation given in Section \ref{S:eqs}, to obtain a
general ``seed'' solution for the relevant B\"acklund transformations.
This allows us in Section \ref{S:sol} to obtain the 1-soliton solution
in a form which suggests a $\tau$-function description. We will
present the $N$-soliton solution of Q3 in Hirota form in
Sec.~\ref{S:NSS}, and show how the solution is related to a set of
discrete Hirota-Miwa equations in a four-dimensional lattice.

\section{The basic lattice equations\label{S:eqs}}
\setcounter{equation}{0} 
The special case $\dd=0$ of (\ref{eq:Q3}),
appeared for first time in \cite{NQC}, (cf. also \cite{QNCL}) in the form
\begin{equation}\label{eq:seq} 
\frac{1+(p-a)s-(p+b)\wt{s}}{1+(q-a)s-(q+b)\wh{s}}=
\frac{1+(q-b)\wt{s}-(q+a)\wh{\wt{s}}}{1+(p-b)\wh{s}-(p+a)\wh{\wt{s}}}\  . 
\end{equation} 
We call this the NQC equation, following \cite{RasinNQC}.

To bring the equation (\ref{eq:seq}) to the form of ${\rm (Q3)_0}$ we
perform the transformation ($a+b\neq 0$):
\begin{equation} \label{eq:us}
u_{n,m}=\tau^n\sg^m\left(s_{n,m}-\tfrac{1}{a+b}\right)\quad,\quad  
\end{equation} 
where
\begin{equation}\label{eq:tausg}
\tau:=\sqrt{\tfrac{(p+a)(p+b)}{(p-a)(p-b)}}\quad,
\quad \sg:=\sqrt{\tfrac{(q+a)(q+b)}{(q-a)(q-b)}}. 
\end{equation}  
This yields ${\rm (Q3)_0}$ with the parametrization
\begin{equation} \label{eq:Q3parm} 
P(u\wh{u}+\wt{u}\wh{\wt{u}})-Q(u\wt{u}+\wh{u}\wh{\wt{u}})
-(p^2-q^2)(\wh{u}\wt{u}+u\wh{\wt{u}})=0,
\end{equation}  
where the lattice parameters have now become points $\ssp=(p,P)$ and 
 $\ssq=(q,Q)$, respectively, on the (Jacobi) elliptic curve:
\begin{equation}\label{eq:parcurves}
P^2=(p^2-a^2)(p^2-b^2)\quad,\quad Q^2=(q^2-a^2)(q^2-b^2). 
\end{equation}  

In this parametrization the $\dd\neq 0$ version of \eqref{eq:Q3} reads
\begin{equation} \label{eq:Q3parmd}
P(u\wh{u}+\wt{u}\wh{\wt{u}})-Q(u\wt{u}+\wh{u}\wh{\wt{u}})
-(p^2-q^2)(\wh{u}\wt{u}+u\wh{\wt{u}})=\dd^2\,\tfrac{(p^2-q^2)}{4PQ}.
\end{equation}

Note that the original parametrization of ${\rm Q3}$ in \eqref{eq:Q3}
is subtly different from \eqref{eq:Q3parm} and they can be related by
the identifications:
\begin{equation} 
\po^2=\tfrac{p^2-b^2}{p^2-a^2},\quad 
\qo^2=\tfrac{q^2-b^2}{q^2-a^2},\qquad 
P=\tfrac{(b^2-a^2)\po}{1-\po^2},\quad  
Q=\tfrac{(b^2-a^2)\qo}{1-\qo^2}. 
\end{equation} 
In fact, the two different parametrizations presented in \eqref{eq:Q3}
and \eqref{eq:Q3parmd} are just different ways to parametrize the
equation
$\alpha(u\wh{u}+\wt{u}\wh{\wt{u}})-\beta(u\wt{u}+\wh{u}\wh{\wt{u}})
-\gamma(\wh{u}\wt{u}+u\wh{\wt{u}})=\dd$, with the constraints arising
from CAC (related to dependency on lattice parameters associated with
different directions of the cube).

It is important for later to observe that the parametrizations of
$P,\,Q,\,\po,\,\qo$ are invariant under the sign change of $a$ and/or
$b$ while the NQC equation itself is not. This means that there are in
fact {\em four different versions} of NQC (corresponding to these sign
changes) and they all provide a different solution to ${\rm (Q3)}_0$,
though the transformation \eqref{eq:us} with corresponding sign
changes in it and \eqref{eq:tausg}. We will use this 4-to-1
correspondence later to construct solutions to Q3.

It was shown in \cite{NQC} that (\ref{eq:seq}) it interpolates through
different choices of the auxiliary parameters $a$, $b$ between various
lattice equations ``of KdV type'', and hence could be thought of as an
interpolating equation. We can identify, e.g., the following subcases
of (\ref{eq:seq}) appearing in the list of \cite{ABS} (up to gauge
transformations, where necessary): $ a=b=0\,\Rightarrow {\rm
  (Q_1)_0}$, $a=0,\,b\rightarrow\infty\,\Rightarrow {\rm (H_3)_0}$ and
$a,b\rightarrow\infty\,\Rightarrow {\rm H_1}$, which are respectively
the lattice Schwarzian KdV equation, the lattice potential modified
KdV equation and the lattice potential KdV equation. For all these
subcases $N$-soliton solutions can be given in closed form, which
follows as an immediate application of the direct linearisation scheme
elaborated in \cite{NQC,QNCL}.

\section{Background and one-soliton solutions\label{S:sol}}
\setcounter{equation}{0} 
\subsection{Seed or background solution for ${\rm Q3}$} 
The trivial solution to \eqref{eq:seq} is $s_{n,m}\equiv 0$ and from
this it follows that $u_{n,m}=c\tau^n\sg^m$ is a background or
``seed'' solution for ${\rm (Q3)_0}$.  Furthermore, as discussed
before, by changing the signs of $a$ and/or $b$ we get other seed
solutions for ${\rm (Q3)_0}$, namely
\begin{equation}\label{eq:sds}
u_{n,m}^{++}=A\tau^n\sg^m,\quad u_{n,m}^{--}=B\tau^{-n}\sg^{-m},\quad
u_{n,m}^{+-}=C\dot\tau^n\dot\sg^m,\quad u_{n,m}^{-+}=D\dot\tau^{-n}\dot\sg^{-m},
\end{equation}
where $\tau$ and $\sg$ were defined in \eqref{eq:tausg} and
\begin{equation}\label{eq:dottausg}
\dot\tau:=\sqrt{\tfrac{(p+a)(p-b)}{(p-a)(p+b)}}\quad,
\quad \dot\sg:=\sqrt{\tfrac{(q+a)(q-b)}{(q-a)(q+b)}}. 
\end{equation}
Starting with one such seed solution for ${\rm (Q3)_0}$ one can use a
Miura transformation (see Appendix) to derive a solution for ${\rm
  (Q3)_\dd}$. The result turns out to be a linear combination of three
of the above terms, and leads us to try a linear combination of all
four terms, that is
\begin{equation}\label{eq:sdcomb}
u_{n,m}^{0SS}=A\tau^n\sg^m+B\tau^{-n}\sg^{-m}
+C\dot\tau^n\dot\sg^m+D\dot\tau^{-n}\dot\sg^{-m}.
\end{equation}
It is easy to verify that this is indeed a solution of ${\rm (Q3)_0}$
provided that
\begin{equation}\label{eq:sdcond0}
AB(a+b)^2-CD(a-b)^2=0,
\end{equation}
and, perhaps surprisingly, that \eqref{eq:sdcomb} is also a solution
of ${\rm (Q3)_\dd}$, provided that
\begin{equation}\label{eq:sdcondd}
AB(a+b)^2-CD(a-b)^2=-\tfrac{\dd^2}{16ab}.
\end{equation}
From this last result we see, that when $\dd\neq 0$ one cannot use any
{\em single} seed given in \eqref{eq:sds} as a starting solution, and
for a ``germinating seed'' (in the terminology of \cite{AHN}) one needs
at least the pair $++,--$ or the pair $+-,\,-+$.

\subsection{1-Soliton solution ${\rm (Q3)}_0$ from BT} 
Having obtained a nontrivial background solution \eqref{eq:sdcomb}, we
can now proceed to construct soliton solutions starting from this seed
solution of the B\"acklund transformation (BT). {}From cubic
consistency if follows that we can impose in addition to the lattice
equation \eqref{eq:Q3parm} also the set of equations:
\bse\label{eq:sides}
\begin{eqnarray}
&& P(u\ol{u}+\wt{u}\wt{\ol{u}})-K(u\wt{u}+\ol{u}\wt{\ol{u}})
=(p^2-k^2)(\wt{u}\ol{u}+u\wt{\ol{u}}+\tfrac{\dd^2}{4PK}),\\
&& Q(u\ol{u}+\wh{u}\wh{\ol{u}})-K(u\wh{u}+\ol{u}\wh{\ol{u}})
=(q^2-k^2)(\wh{u}\ol{u}+u\wh{\ol{u}}+\tfrac{\dd^2}{4QK}),
\end{eqnarray}
\ese
where $K^2=(k^2-a^2)(k^2-b^2)$.  Since the ``bar''-direction is for
increasing number of solitons, we search for a new $\ol{u}(\equiv
u^{1SS})$ of the form: $\ol{u}=\ol{u}_\theta+v$, where $\ol{u}_\theta$
is the bar-shifted background solution \eqref{eq:sdcomb}:
\begin{equation}\label{eq:sdb}
\bar u_\theta=A\tau^n\sg^m\kappa+B\tau^{-n}\sg^{-m}\kappa^{-1}
+C\dot\tau^n\dot\sg^m\dot\kappa+D\dot\tau^{-n}\dot\sg^{-m}\dot\kappa^{-1},
\end{equation}
where
\begin{equation}\label{eq:kappa}
\kappa=\sqrt{\tfrac{(k+a)(k+b)}{(k-a)(k-b)}},\quad
\dot\kappa=\sqrt{\tfrac{(k+a)(k-b)}{(k-a)(k+b)}}.
\end{equation}
(Note however, that we can absorb the $\kappa,\dot\kappa$ factors into
$A,B,C,D$ without changing the relation \eqref{eq:sdcond0} or
\eqref{eq:sdcondd}.)

From \eqref{eq:sides} one can then solve
\begin{equation} \wt{v}=\frac{ 
[(p^2-k^2)\wt{u}_\theta+K\wt{\ol{u}}_\theta)-Pu_\theta]v}
{ -Kv +[P\wt{u}_\theta-K\ol{u}_\theta-(p^2-k^2)u_\theta]},\, 
\wh{v}=\frac{ [(q^2-k^2)\wh{u}_\theta+K\wh{\ol{u}}_\theta)-Qu_\theta]v}
{ -Kv +[Q\wh{u}_\theta-K\ol{u}_\theta-(q^2-k^2)u_\theta]}
\end{equation}
where the expected $\dd^2$ term in the numerator disappears by virtue
of the definitions (\ref{eq:sdb},\ref{eq:sdcondd}). Then introducing
$v=f/g$ we can linearise these equations and obtain
\begin{equation}
  \wt{\bphi}=\Lambda \left(\begin{array}{cc} 
(p^2-k^2)\wt{u}_\theta+K\wt{\ol{u}}_\theta-Pu_\theta & 0 \\  
      -K & P\wt{u}_\theta-K\ol{u}_\theta-(p^2-k^2)u_\theta 
\end{array}\right)\,\bphi 
\end{equation} 
and similarly for $\wh{\bphi}$, where $\bphi=(f,g)^T$. The factor
$\Lambda$ is determined by the condition that the shifts with
$\wt{\phantom{a}}$ and $\wh{\phantom{a}}$ must commute, from which it
follows we should take $\Lambda\propto 1/U_\theta$, where
(compare with \eqref{eq:sdcomb}) 
\begin{equation}\label{eq:Uth}
(U_\theta)_{nm}=(a+b)A\tau^n\sg^m-(a+b)B\tau^{-n}\sg^{-m}
+(a-b)C\dot\tau^n\dot\sg^m-(a-b)D\dot\tau^{-n}\dot\sg^{-m}.
\end{equation}

Then with some algebra we obtain:
\begin{equation} \label{eq:phisol}
\bphi_{n,m} = (p-k)^n(q-k)^m
\left(\begin{array}{ccc} 
\rho_{n,m} (U_\theta)_{n,m}/(U_\theta)_{0,0}  &,& 0 \\  
 \frac{K}{2k}[1-\rho_{n,m}]/(U_\theta)_{0,0} &,& 1
\end{array}\right)\,\bphi_{0,0}  \, , 
\end{equation} 
where $\rho_{n,m}$ is the discrete  "plane-wave factor", defined by
\begin{equation}\label{eq:rho}
\rho_{n,m}=\left(\frac{p+k}{p-k}\right)^n\left(\frac{q+k}{q-k}\right)^m\,
\rho_{0,0}.
\end{equation}
{}From the result \eqref{eq:phisol} we can reconstruct $v$ of the 1SS:
\begin{equation}\label{eq:1ss}
  v_{n,m}=\frac{2k(U_\theta)_{nm}\rho_{nm}v_{00}}
{2k(U_{\theta})_{00}+K v_{00}(1-\rho_{nm})},
\end{equation}
and finally $u_{nm}^{1SS}=\bar u_\theta+v$. 

For a more explicit form showing the $A,B,C,D$ dependence, we redefine
the constant $\rho_{0,0}$ in \eqref{eq:rho} so that the denominator
becomes proportional to $1+\rho_{nm}$. Furthermore, after scaling
$A,B,C,D$ so that $\bar u_\theta$ of \eqref{eq:sdb} becomes $u^{0SS}$
of \eqref{eq:sdcomb} we can write the 1SS \eqref{eq:1ss} as
\begin{align}
  u^{1SS}_{nm}=&\left[
 A\tau^n\sg^m \left(1+\rho_{nm}\tfrac{(a-k)(b-k)}{(a+k)(b+k)}\right)
+B\tau^{-n}\sg^{-m}\left(1+\rho_{nm}\tfrac{(a+k)(b+k)}{(a-k)(b-k)}\right)
\right.\nonumber\\
&+\left.C\dot\tau^n\dot\sg^m\left(1+\rho_{nm}
\tfrac{(a-k)(b+k)}{(a+k)(b-k)}\right)
+D\dot\tau^{-n}\dot\sg^{-m}\left(1+\rho_{nm}
\tfrac{(a+k)(b-k)}{(a-k)(b+k)}\right)
\right]/(1+\rho_{nm}).
  \label{eq:1SS}
\end{align}

\section{$N$-soliton solutions and Hirota bilinear form\label{S:NSS} }
\setcounter{equation}{0} We will now present the main result of the
paper, which is a general $N$-soliton solution of Q3. This solution
can be written in the following form:
\begin{align}
  u_{nm}=&
 A\tau^n\sg^m\,\frac{F(n,m,\alpha+1,\beta+1)}{F(n,m,\alpha,\beta)}
+B\tau^{-n}\sg^{-m}\,
\frac{F(n,m,\alpha-1,\beta-1)}{F(n,m,\alpha,\beta)}
\nonumber\\
&+C\dot\tau^n\dot\sg^m\, \frac{F(n,m,\alpha+1,\beta-1)}{F(n,m,\alpha,\beta)}
+D\dot\tau^{-n}\dot\sg^{-m}\, \frac{F(n,m,\alpha-1,\beta+1)}
{F(n,m,\alpha,\beta)},
  \label{eq:1SSF}
\end{align}
where the tau-function $F$ is given by
\begin{equation}
  \label{eq:FNSS}
  F(n,m,\alpha,\beta )=\sum_{\mu_j\in\{0,1\}}
\exp\left(\sum_{j=1}^N\mu_j\,\eta_j+\sum_{1\le i<j\le
      N}\mu_i\mu_ja_{ij}\right),
\end{equation}
where
\begin{equation}
  \label{eq:rhodef}
 \exp{\eta_j}\equiv \rho_{nm\alpha\beta}(k_j):=\left(\frac{p+k_j}{p-k_j}\right)^n
\left(\frac{q+k_j}{q-k_j}\right)^m
\left(\frac{a-k_j}{a+k_j}\right)^\alpha
\left(\frac{b-k_j}{b+k_j}\right)^\beta\rho_0,
\end{equation}
\begin{equation}
  \label{eq:FAdef}
\exp\, a_{ij}\equiv A_{ij}:=\left(\frac{k_i-k_j}{k_i+k_j}\right)^2,
\end{equation}
and the coefficients in \eqref{eq:1SSF} are restricted by
\begin{equation}\label{eq:sdcondd2}
AB(a+b)^2-CD(a-b)^2=-\tfrac{\dd^2}{16ab}.
\end{equation}

One way in which this result can be derived is by using the Miura
transformation given in the Appendix and using a number of relations
following from the direct linearisation structure of the NQC equation
of \cite{NQC,QNCL}. We will not exhibit this approach here for lack of
space. Rather, we will argue the result based on the connection with
the Hirota-Miwa difference equations, \cite{Hirota,Miwa}.

Having established the 1SS \eqref{eq:1SS}, which is special case of
the form \eqref{eq:1SSF} with \eqref{eq:FNSS}, we checked the 2SS
suggested by Hirota's perturbative approach
\begin{equation}
  \label{eq:F2SS}
F(n,m,\alpha,\beta )=1+\rho_{nm\alpha\beta}(k_1)+\rho_{nm\alpha\beta}(k_2)
+A_{12}\rho_{nm\alpha\beta}(k_1)\rho_{nm\alpha\beta}(k_2).
\end{equation}
and verified that the phase factor $A_{ij}$ is given as in
\eqref{eq:FAdef}. We have also explicitly verified that the 3SS
version solves Q3.

The form of the phase factor \eqref{eq:FAdef} allows us to identify
the solution to be in the Hirota-Miwa hierarchy \cite{Hirota,Miwa},
within the reduction $-q_i=p_i\equiv k_i$ (up to some notational
conventions for the parameters), except that we now seem to have a 4D
system $n,m,\alpha,\beta$ with diagonal [$(\alpha\pm 1,\beta\pm 1)$]
and anti-diagonal [$(\alpha\pm 1,\beta\mp 1)$] reductions.

The $N$-soliton $F$ in \eqref{eq:FNSS} solves the Hirota-Miwa bilinear
difference equations in any 3 of the 4 indices (along with some higher
equations in the hierarchy), e.g., \bse \label{eq:HM4}
\begin{align}
   (q+a) F_{1} F_{23}
& -(a+p) F_{2} F_{13} +(p-q) F_{12} F_{3}=0,\label{he1}\\
(q+b) F_{1} F_{24}
& -(b+p) F_{2} F_{14} +(p-q) F_{12} F_{4}=0,\label{he7}\\
 (p - q)F_{123} F_{}
&  + (q -a) F_{13} F_{2}  - (p -a)F_{1} F_{23} =0,\label{he21}\\
(p - q)F_{124} F_{} 
&  + (q -b) F_{14} F_{2}  - (p -b) F_{1} F_{24} =0.\label{he27}
\end{align}
\ese 
Here we have used the notation where only the shifted index is
indicated, e.g., $F_{n,m+1,\alpha+1,\beta+1}\equiv F_{234}$.  The
shift in the negative direction is indicated by a bar, e.g.,
$F_{n,m,\alpha-1,\beta+1}\equiv F_{\bar 34}$.  In the following we
also need the various shifts of these equations. (The equations above
would be more symmetric if we were to change $a\to -a$ $b\to -b$, cf.
\eqref{eq:rhodef}.)

Using the above equations we can show that \eqref{eq:1SSF} solves
\eqref{eq:Q3parm}. After substitution we collect terms with different
factors of $\tau^n,\,\sigma^m,\,\dot\tau^n,\,\dot\sigma^m$ or
equivalently different combinations of the coefficients $A,B,C,D$,
except that $AB,\, CD$ and $\dd^2$ should go together. For example the
coefficient of $D^2$ (or $\dot\tau^{-2n}\dot\sigma^{-2m}$) is quartic
in the $F$'s, but it can be decomposed into various combinations of
Hirota-Miwa equations \eqref{eq:HM4} as follows
\begin{align*}
F_{2\bar 3}\times(\text{coeff. of }D^2)\propto\,
&[\eqref{he21}_{\bar3} F_{12\bar34}-\eqref{he27}_{\bar3} F_{12}]
[(a+q) F_{2} F_{\bar 34} - (b+q) F F_{2\bar34}]\\
&-[\eqref{he1}_{\bar3} F_{\bar 34}-\eqref{he7}_{\bar3} F)] 
[(a-p) F_{2} F_{12\bar34} - (b-p) F_{12} F_{2\bar34}].
\end{align*}
and similarly for the coefficients of $A^2,\,B^2,\,C^2$. The
coefficient of $AC$, on the other hand, is proportional to
\begin{align*}
&\{[-\eqref{he7}_3 F+\eqref{he21} F_{34}] 
[(a+p) F_2 F_{123\bar4}-(b+p) F_{12} F_{23\bar4}]\\
&\quad+[\eqref{he7}_{3\bar4} F_{12}-\eqref{he1} F_{123\bar4}] 
[(a-q) F_2 F_{34}+(b+q) F F_{234}]\} (b-p) (b-q)\\
-&\{[\eqref{he27}_{3\bar4} F-\eqref{he21} F_{3\bar4}] 
[(a+p) F_2 F_{1234}+(b-p) F_{12} F_{234}]\\
&\quad -[\eqref{he27}_3 F_{12}-\eqref{he1} F_{1234}] 
[(a-q) F_2  F_{3\bar4}-(b-q) F F_{23\bar4}]\} (b+p) (b+q)
\end{align*}
and similar formulae can be written for the coefficients of
$AD,\,BC,\,BD$. 

This leaves the equation containing $AB,\,CD$, and $\dd^2$, which can
be written as
\begin{equation}
  \label{eq:ABc1}
AB\mathcal R(b)+CD\mathcal R(-b)+\dd^2\mathcal S=0  
\end{equation}
where $\mathcal R$ and $\mathcal S$ are some expression in $F$ and the
parameters and $\mathcal S$ does not depend on $b$. The reflection
property with respect to $b$ is obvious when one considers the role of
$b$ in (\ref{eq:tausg}, \ref{eq:dottausg}, \ref{eq:1SSF},
\ref{eq:rhodef}).

One of the fundamental assumptions before was that although the coefficients
$A,B,C,D$ may depend on $\dd$, the soliton part $F$ does not. Thus
equation \eqref{eq:ABc1} should hold under \eqref{eq:sdcondd2},
whether or not $\dd=0$.  If $\dd=0$ we get immediately the condition
\begin{equation}
  \label{eq:ABc2}
(a+b)^2\mathcal R(b)+(a-b)^2\mathcal R(-b)=0.
\end{equation}
and then using this we get for $\dd\neq0$
\begin{equation}
  \label{eq:ABc3}
\mathcal R(b)+16ab(a-b)^2\mathcal S=0.
\end{equation}
If \eqref{eq:ABc3} holds, so does its $b\to -b$ reflection, and they
together imply \eqref{eq:ABc2}. In full detail equation
\eqref{eq:ABc3} reads
\begin{align*}
F&_{12\bar34} F_{13\bar4} F_{2} F (a+p) (a-p) (a-q) (b+p) (b-p) (b+q)\\
&-F_{12\bar34} F_1 F_{23\bar4} F (a-p) (a+q) (a-q) (b+p) (b+q) (b-q)\\
&+F_{12\bar34} F_1 F_{2} F_{3\bar4} (a-p) (a-q) (b+p) (b+q) (p+q) (p-q)\\
&+F_{123\bar4} F_{1\bar34} F_{2} F (a+p) (a-p) (a+q) (b+p) (b-p) (b-q)\\
&-F_{123\bar4} F_1 F_{2\bar34} F (a+p) (a+q) (a-q) (b-p) (b+q) (b-q)\\
&+F_{123\bar4} F_1 F_{2} F_{\bar34} (a+p) (a+q) (b-p) (b-q) (p+q) (p-q)\\
&+F_{12} F_{1\bar34} F_{23\bar4} F (a-p) (a+q) (b+p) (b-q) (p+q) (p-q)\\
&-F_{12} F_{1\bar34} F_{2} F_{3\bar4} (a-p) (a+q) (a-q) (b+p) (b+q) (b-q)\\
&+F_{12} F_{13\bar4} F_{2\bar34} F (a+p) (a-q) (b-p) (b+q) (p+q) (p-q)\\
&-F_{12} F_{13\bar4} F_{2} F_{\bar34} (a+p) (a+q) (a-q) (b-p) (b+q) (b-q)\\
&+F_{12} F_1 F_{2\bar34} F_{3\bar4} (a+p) (a-p) (a-q) (b+p) (b-p) (b+q)\\
&+F_{12} F_1 F_{23\bar4} F_{\bar34} (a+p) (a-p) (a+q) (b+p) (b-p) (b-q)\\
&+4 F_{12} F_1 F_{2} F (a-b)^2 (p+q) (p-q) a b=0,
\end{align*}
and it must also be expressible in terms of the equations in the
Hirota-Miwa hierarchy.  (We have verified that this equation is solved
by the 3SS of the type \eqref{eq:FNSS}.)

\section{Conclusions}
In this paper we have constructed NSS to Q3. This was done through an
associated equation, the NQC-equation, which contains extra parameters
$a,b$. The correspondence from NQC to ${\rm (Q3)}_{0}$ is 4-to-1,
labeled by the different sign combinations of $a,b,$ and this leads us
to a four-term background solution, from which the 1SS and then the
NSS can be constructed. These solutions contain the new parameters
$a,b$ in an essential manner. An important observation is the
relationship with the Hirota-Miwa equation with four variables.

Another proof based on the known NSS of the NQC equation, as well as
other details and properties will be given elsewhere \cite{AN2}.

\ack One of the authors (JH) would like to that JJC Nimmo for
discussions.  JA was supported by the UK Engineering and Physical
Sciences Research Council.

\section*{Appendix: Miura transformation between  ${\rm (Q3)}_0$ 
and ${\rm (Q3)}_\delta$}
\def\theequation{A.\arabic{equation}} \setcounter{equation}{0}
\def\thesubsection{A.\arabic{subsection}}
In the Section \ref{S:sol} we constructed a 1SS \eqref{eq:1SS} starting
with the most general 0SS \eqref{eq:sdcomb} and using the CAC-cube, in
which the vertical direction associated with the B\"acklund
transformation.  Here we again use the CAC-cube, but now the bottom
layer is ${\rm (Q3)}_\delta$, the top layer ${\rm (Q3)}_0$, with the
sides providing linear equations by which a solution of ${\rm (Q3)}_0$
can be transformed into a solution of ${\rm (Q3)}_\dd$.  For example
the well known NSS, corresponding to $A=1,\,B=C=D=0$, which satisfies
\eqref{eq:sdcond0} but not \eqref{eq:sdcondd}, can be transformed into
a solution of ${\rm (Q3)}_\dd$ and therefore this transformation
should generate at least a $B$ term \cite{AN2}.

\subsection{Derivation of Miura transformation} 
In order to get the $\dd=0$ version of (\ref{eq:Q3parmd}) on the top
layer of the cube we use a simple scaling argument. Let us again
denote the vertical shift by a bar and introduce the scaling
$\ol{u}=v/\epsilon$, with $\epsilon\rightarrow 0$, where $v$ solves
${\rm (Q3)}_0$. In this limit the top equation immediately becomes
\eqref{eq:Q3parm}. For the side equation we get first
\[ 
P(uv+\wt{u}\wt{v})-\epsilon
R(u\wt{u}+\tfrac{1}{\epsilon^2}v\wt{v})=(p^2-r^2)\left(
  u\wt{v}+\wt{u}v+ \tfrac{\epsilon\dd^2}{4PR}\right),
\]
and a similar equation for the shift in the other lattice direction
associated with $q$. Arranging the parameter $R$ to be of order
$\epsilon$, i.e. $R=\epsilon R_0$ with $R_0$ finite and nonzero as
$\epsilon\rightarrow 0$, yields in dominant order in $\epsilon$ the
equation
\[ 
P(uv+\wt{u}\wt{v})- R_0 v\wt{v}=(p^2-r^2)\left( u\wt{v}+\wt{u}v+
\tfrac{\dd^2}{4PR_0}\right).
\]
However, we must respect the parametrization of \eqref{eq:parcurves},
which now reads
\[ 
\epsilon^2 R_0^2=(r^2-a^2)(r^2-b^2),
\] 
and this can only be of the right order in $\epsilon$ if
~$r^2=a^2+\epsilon^2 r_0^2$~, or ~$r^2=b^2-\epsilon^2 r_0^2$~ with
$r_0$ finite as $\epsilon\rightarrow 0$. Hence,
~$R_0^2=r_0^2(a^2-b^2)$~, but since $r_0$ is arbitrary, we may keep
$R_0$ arbitrary and only remember that either $r^2=a^2$ or $r^2=b^2$.
Thus, we are led to the following Miura transformation between $({\rm
  Q3})_\delta$ and ${\rm (Q3)}_0$:
\bse\label{eq:Miura}\begin{eqnarray}\label{eq:Miuraa}
  P(uv+\wt{u}\wt{v})-Rv\wt{v}=(p^2-r^2)\left(\wt{u}v+u\wt{v}
+\tfrac{\dd^2}{4PR}\right), \\
  Q(uv+\wh{u}\wh{v})-Rv\wh{v}=(q^2-r^2)\left(\wh{u}v+u\wh{v}
+\tfrac{\dd^2}{4QR}\right),
\label{eq:Miurab}  
\end{eqnarray}\ese  
where we have suppressed the suffix 0.

In (\ref{eq:Miura}) $v$ is the solution of ${\rm (Q3)}_0$ and $u$ a
corresponding solution of ${\rm (Q3)}_\delta$, and we note that the
resulting equations are actually \textit{linear} in the new solution
$u$: 
\bse\label{eq:Lin}\begin{eqnarray}\label{eq:Lina}
  && [Pv-(p^2-a^2)\wt{v}]u+ [P\wt{v}-(p^2-a^2)v]\wt{u}=
Rv\wt{v}+\tfrac{\dd^2(p^2-a^2)}{4PR}\  , \\
  && [Qv-(q^2-a^2)\wh{v}]u+
  [Q\wh{v}-(q^2-a^2)v]\wh{u}=Rv\wh{v}+\tfrac{\dd^2(q^2-a^2)}{4QR}\ ,
\label{eq:Linb}  
\end{eqnarray}
\ese where we have chosen $r^2=a^2$.  It can be shown that the Miura
transformation (\ref{eq:Lin}) commute on the two-dimensional lattice, as a
consequence of which we can ``integrate'' the linear relations
(\ref{eq:Lina}) and (\ref{eq:Linb}) simultaneously in both lattice
directions to obtain a well-defined solution, since the lattice
equation obeyed by $v$ is a member of the equations consistent around
the cube. 

Other Miura type transformations between various members of the ABS
list of lattice equations, and related equations that can be obtained
by degeneration have been constructed, and will be published
separately \cite{Atkinson}.

\subsection{From seed of $({\rm Q3})_0$ to the seed of $({\rm Q3})_\delta$}
We now apply the Miura transformation (\ref{eq:Miura}) starting with
$v$ given by the background solution ~$v_{n,m}=-\tau^n\sg^m/(a+b)$.
In that case the coefficients of the linear relation (\ref{eq:Lina})
are given by:
\begin{equation}\label{eq:seedcoeffs} 
Pv-(p^2-a^2)\wt{v}=\tfrac{P}{p-b}\tau^n\sg^m\quad,\quad 
P\wt{v}-(p^2-a^2)v=-\tfrac{P}{p+b}\tau^{n+1}\sg^m\  , 
\end{equation} 
and similarly for the coefficients of (\ref{eq:Linb}) interchanging
the roles of $\ssp$ and $\ssq$ and of $n$ and $m$.  Inserting these
coefficients we obtain the linear equations
\bse\label{eq:Lineqseed}\begin{eqnarray}
  u-\tfrac{p-b}{p+b}\,\tau\,\wt{u}&=&
  \tfrac{R(p-b)}{P(a+b)^2}\,\tau^{n+1}\sg^m
+\tfrac{\delta^2}{4(p+b)R}\,\tau^{-n}\sg^{-m}, \\
  u-\tfrac{q-b}{q+b}\,\sg\,\wh{u}&=&
  \tfrac{R(q-b)}{Q(a+b)^2}\,\tau^{n}\sg^{m+1}
+\tfrac{\delta^2}{4(q+b)R}\,\tau^{-n}\sg^{-m}.
\end{eqnarray} \ese 
The first relation can be integrated w.r.t.~the variable $n$,
using  $\dtau^n$ defined in \eqref{eq:dottausg} as integrating factor,
and we obtain
\[ 
u_{0,m}-\dtau^n u_{n,m}= \tfrac{R\sg^m}{2a(a+b)^2}
\left[ \bigl(\tfrac{p+a}{p-a}\bigr)^n-1\right]
-\tfrac{\dd^2\sg^{-m}}{8bR}\left[
  \bigl(\tfrac{p-b}{p+b}\bigr)^n-1\right]\  ,
\] 
What is significant here is that the coefficients on the right hand
side no longer depend on the lattice parameter $p$, and that $p$ and
$q$ only occur through the discrete ``exponentials''. Obviously a
similar relation can be obtained from
(\ref{eq:Lineqseed}b) integrating w.r.t. variable $m$, and the result
is given by simply interchanging $p$ ad $q$ and the roles of $n$ and
$m$;
\[ 
u_{n,0}-\dsg^m u_{n,m}= \tfrac{R\tau^n}{2a(a+b)^2}
\left[ \left(\tfrac{q+a}{q-a}\right)^m-1\right]
-\tfrac{\dd^2\tau^{-n}}{8bR}\left[
  \left(\tfrac{q-b}{q+b}\right)^m-1\right]\  ,
\] 
with $\dsg$ defined in \eqref{eq:tausg}.
 
Now we can either take the first result and add a multiple of the
second result at $n=0$, or take the second result and add a multiple
of the first result at $m=0$, thus eliminating the intermediate terms
$u_{0,m}$ and $u_{n,0}$ respectively.  In either way we get the
following result:
\[
u_{nm}=-\tau^n\sigma^m\tfrac{R}{2a(a+b)^2}
+\tau^{-n}\sigma^{-m}\tfrac{\dd^2}{8bR}
+\dot\tau^{-n}\dot\sigma^{-m}(u_{00}-\tfrac{\dd^2}{8bR}+\tfrac{R}{2a(a+b)^2}).
\]
Thus, starting with the background solution for  $({\rm Q3})_0$ with
$A=1,B=C=D=0$ we generated a solution of  $({\rm Q3})_\delta$ with
$A=-{R}/({2a(a+b)^2}),\,B=\dd^2/(8bR),\,C=0,\,D\neq 0$, which indeed
satisfies the constraint \eqref{eq:sdcondd}.

This three-term solution can be suggestively written in as follows:
\[ 
u_{n,m}= A\tau^n\sg^m+C\dtau^{-n}\dsg^{-m}+D\tau^{-n}\sg^{-m}\   ,
\] 
which contains all but one of the possible sign interchangements of
the parameters $a$ and $b$ in the discrete exponential factors.  This
seems to suggest that there is one term missing, which could have been
obtained by using a Miura transformation with $r=\pm b$, instead of
$r=\pm a$ as we have done. Thus we reach the ansatz \eqref{eq:sdcomb}.

\end{document}